\documentclass[11pt]{article}
\usepackage{epsfig,amsmath,amssymb,stmaryrd,enumerate}
\usepackage{graphicx}

\usepackage{latexsym}
\usepackage{amsthm}
\usepackage{amsmath}
\usepackage{amsfonts, epsfig, amsmath, amssymb, color, amscd}
\usepackage[cmtip,arrow]{xy}
\usepackage{pb-diagram, pb-xy}
\usepackage{amssymb,epsfig}
\usepackage{color}
\usepackage[cmtip,arrow]{xy}
\usepackage{pb-diagram, pb-xy}
\usepackage{amssymb,epsfig,amsfonts}

\newfont{\sdbl}{msbm9}
\newfont{\dbl}{msbm10 at 12pt}
\theoremstyle{definition}

\theoremstyle{plain}

\newtheorem{theorem}{Theorem}

\begin{document}

\title{On commuting ordinary differential operators with polynomial coefficients corresponding to spectral curves of genus two
\thanks{The authors were supported by the Russian Foundation for Basic Research  (grant 16-51-55012). The second author (A.E.M.) was also supported by a grant from Dmitri Zimin's
''Dynasty'' foundation.}}

\author{{Valentina N. Davletshina,\ Andrey E. Mironov }}

\date{}
\maketitle

\begin{abstract}
The group of automorphisms of the first Weyl algebra acts on commuting ordinary differential operators with polynomial coefficient. In this paper we prove that for fixed generic spectral curve of genus two the set of orbits is infinite.
\end{abstract}

\section{Introduction and Main Result}

Let us consider a generic equation
\begin{equation*}\label{1}
f(X,Y)=\sum_{i,j}\alpha_{ij}X^iY^j=0,\qquad  \alpha_{ij}\in{\mathbb C}.
\end{equation*}
The group $\it{Aut(A_1)}$, where  $A_1=\mathbb{C}[x][\partial_x]$ is the first Weyl algebra, has a natural action on the set of solutions of this equation. Yu.~Berest proposed the following conjecture (cf. \cite{MZh}):  if the genus of the algebraic curve defined  by the equation $f(z,w)=0$ (in the theory of commuting ordinary differential operators the curve defined by this equation is called {\it the spectral curve}) is one then the set of orbits
is infinite, and if the genus is greater than one then for generic $\alpha_{ij}$ the set of orbits is finite.
From finiteness of this set for some curve it would be possible to derive the Dixmier
conjecture ${\it End(A_1)}=\it{Aut(A_1)}$ for the first Weyl algebra.

In this paper we study the action of the automorphisms group of  the first Weyl algebra $A_1=\mathbb{C}[x][\partial_x]$ on the set of solutions of the equation
\begin{equation}
\label{11}
Y^2=X^{2g+1}+c_{2g}X^{2g}+\ldots+c_1X+c_0
\end{equation}
at $g=2$.
It is not difficult to show that any solution $X,Y$ of this equation is a pair of commuting operators.
 To construct examples of commuting operators from $A_1$ satisfying \eqref{11} is a nontrivial problem. First such examples, for $g=1$,  were found by Dixmier in \cite{Dix}.
For $g>1$ the examples  of higher rank commuting ordinary differential operators were constructed, using Krichever--Novikov theory (see \cite{K1}, \cite{KN}), in \cite{M}: the
operator
$$
 L_4=(\partial_x^2+\alpha_3x^3+\alpha_2x^2+\alpha_1x+\alpha_0)^2+g(g+1)\alpha_3x
$$
commutes with an operator $L_{4g+2}\in A_1$ of order $4g+2$ and $L_4,L_{4g+2}$ satisfy \eqref{11} (another examples see in \cite{D}--\cite{Og2}).

In \cite{MZh} the set of orbits was studied in the case of genus one spectral curves, i.e. for $g=1$ in \eqref{11}.
It was shown that the set of orbits is infinite for any such curve.
Moreover, for arbitrary $g>1$ there is a two-parametric family of hyperelliptic spectral curves with infinite set of orbits \cite{MZh}.
 More precisely, the operator
$$
L_4^{\sharp}=(\partial_x^2+\alpha_1 cosh(x)+\alpha_0)^2+\alpha_1 g(g+1)cosh(x), \ \ \ \ \alpha_1\neq0
$$
commutes with an operator $L_{4g+2}^\sharp$ of order $4g+2$ (see \cite{M1}), and
the pair $L_{4}^\sharp,$ $L_{4g+2}^\sharp$ satisfies the equation
\begin{equation}
\label{2}
\Big(L_{4g+2}^\sharp\Big)^2=\Big(L_{4}^\sharp\Big)^{2g+1}+c_{2g}^\sharp \Big(L_{4}^\sharp\Big)^{2g}+\ldots+c_{1}^\sharp L_{4}^\sharp+c_0^\sharp
\end{equation}
for some constants $c_{i}^\sharp$. Mokhov \cite{Mokh} noticed that the change of variables
$$
x=ln(y+\sqrt{y^2-1} )^r, \ \ \ \ \ \ r=\pm1, \ \pm2,\ldots,
$$
transforms the operators $L_{4}^\sharp,$ $L_{4g+2}^\sharp$ into the operators with polynomial coefficients in new variable $y,$ e.g.
\begin{equation*}
\label{3}
L_{4}^\sharp=((1-y^2)\partial_y^2-3y\partial_y+aT_r(y)+b)^2-ar^2g(g+1)T_r(y), \ \ \ a\neq0.
\end{equation*}
Here $T_r(y)$ is the Chebyshev polynomial of degree $|r|$. Recall that
$$
T_0(y)=1, \ \ T_1(y)=y, \ \ T_r(y)=2yT_{r-1}(y)-T_{r-2}(y), \ \ T_{-r}(y)=T_r(y).
$$
This gives a family of operators $L_{4}^\sharp,$ $L_{4g+2}^\sharp\in A_1$ depending on integer $r$ which satisfy \eqref{2}.
It is turn out that for different integers $r$ the pairs $L_{4}^\sharp,$ $L_{4g+2}^\sharp$ belong to different orbits (see \cite{MZh}), so for the equation
$$
 Y^2=X^{2g+1}+c_{2g}^\sharp X^{2g}+\ldots+c_{1}^\sharp X+c_0^\sharp
$$
the set of orbits is infinite.

The main result of this paper is the following. We give a simple family (probably, the simplest one) of commuting operators from $A_1$, satisfying the equation
\begin{equation}
\label{th}
 Y^2=X^5+c_4X^4+c_3X^3+c_2X^2+c_1X+c_0, \ \ \ X,Y\in A_1, \ \ c_i\in \mathbb{C}
\end{equation}
with $c_i$ being generic,  which belong to different orbits. The proof of the last fact  appears to be also extremely simple for this family.

\begin{theorem}
The operator
$$
 L^{^\flat}_4=((\alpha_1x^2+1)\partial_x^2+(\alpha_2x+\alpha_3)\partial_x+\alpha_4x+\alpha_5)^2+\alpha_1\alpha_4g(g+1)x+\alpha_6
$$
commutes with an operator $L^{^\flat}_{10}$ (given by exact formulae in Appendix) at $g=2$ for any $\alpha_i\in \mathbb{C}$.

The pair $L^{^\flat}_4, L^{^\flat}_{10}$ is a solution of \eqref{th}, where $c_i$ depend polynomially on $\alpha_i$ (see exact formulae in Appendix).

The set of orbits of the group $\it{Aut(A_1)}$ in the space of solutions of the equation \eqref{th} with generic $c_i$
is infinite.
\end{theorem}

We also checked that $L^{^\flat}_4$ commutes with $L^{^\flat}_6$ at $g=1$ and with $L^{^\flat}_{14}$ at $g=3.$ So
we can formulate the conjecture:

\vskip5mm

{\it The operator $L^{^\flat}_{4}$ commutes with an operator of order $4g+2$}.

\vskip5mm

The operator $L^{^\flat}_4$ is not self-adjoint hence the methods of \cite{M} are not applicable here. It is an interesting problem to prove this conjecture and develop the methods
of \cite{M} to the non-self-adjoint case.

The authors are sincerely grateful to Alexander Zheglov for valuable discussion and comments.

\section{Proof of Theorem 1}

The theorem follows from the following observation. Direct calculations show that the operator
$$
 L^{^\flat}_4=((\alpha_1x^2+1)\partial_x^2+(\alpha_2x+\alpha_3)\partial_x+\alpha_4x+\alpha_5)^2+\alpha_1\alpha_4g(g+1)x+\alpha_6
$$
commutes with an operator $L^{^\flat}_{10}$ at $g=2$. Operators $L^{^\flat}_4,L^{^\flat}_{10}$ satisfy \eqref{th}, where the coefficients $c_i$ depend polynomially on $\alpha_i$ (see exact formulae in Appendix). Thus, we obtain an algebraic morphism ${\mathbb{C}}^6 \rightarrow {\mathbb{C}}^5$. Direct calculations show that the differential map (on tangent spaces) at a generic point is surjective, thus our morphism  is dominant.
Hence, if we fix generic coefficients $c_j$, we obtain 1-parametric family
of solutions of \eqref{th}. Let's show that operators from this family belong to infinitely many different orbits of the group $\it{Aut(A_1)}$.

Recall that $\it{Aut(A_1)}$ is generated by the following automorphisms (see \cite{Dix})

$$\varphi_1(x)=x+P_1(\partial_x), \ \ \ \ \varphi_1(\partial_x)=\partial_x,$$
$$\varphi_2(x)=x, \ \ \ \ \ \ \ \varphi_2(\partial_x)=\partial_x+P_2(x),$$
$$\varphi_3(x)=\alpha x+\beta \partial_x, \ \ \ \ \varphi_3(\partial_x)=\gamma \partial_x+\delta x, \ \ \ \alpha\gamma-\beta\delta=1, \ \ \ \alpha, \beta, \gamma, \delta\in \mathbb{C},
$$
where $P_1,P_2$ are arbitrary polynomials with constant coefficients.

 Let $S(L)$ denote the total symbol of an operator $L\in A_1$ (thus, $S(L)$ is a polynomial in two variables; we assume here  that operators are written in a canonical form, say, with coefficients in $x$ on the left). Let's remind two simple properties of the total symbol: $S(L_1+L_2)= S(L_1)+S(L_2)$ and
$\deg (S(L_1L_2))= \deg (S(L_1)) + \deg (S(L_2))$ for any $L_1,L_2\in A_1$.

Without loss of generality we can assume that $\alpha_1\neq 0$, for if $\alpha_1= 0$, then the coefficients $c_i$ form an algebraic subset of dimension one in ${\mathbb{C}}^5$ (as it follows from the exact formulae in Appendix).

Now let $\varphi \in Aut(A_1)$. Consider two possibilities: either $S(\varphi(x))$ or $S(\varphi(\partial_x))$ has degree $>$ 1, or $\deg S(\varphi(x))=\deg S(\varphi(\partial_x))=1$.

In the first case, as it obviously follows from the two simple properties of the total symbol and the standard property of the polynomial degree:
$$\deg (S(\varphi (L^{^\flat}_4)))=4(\deg S(\varphi(x)) + \deg S(\varphi(\partial_x)))> 8=\deg (S(L^{^\flat}_4)),
$$
a contradiction.

In the second case
$$
 \varphi(x)=\alpha x+\beta \partial_x +C_1,\qquad
 \varphi(\partial_x)=\gamma \partial_x+\delta x +C_2,
$$
where $\alpha\gamma-\beta\delta=1$, $\alpha, \beta, \gamma, \delta , C_1, C_2\in \mathbb{C}$, and an easy direct calculation shows that such an automorphism preserves the form of the operator $L^{^\flat}_4$ only if $C_1=C_2=\beta =\delta =0$, $\alpha =\gamma =1$.

Thus, the operators $L^{^\flat}_4$ belong to different orbits for different values of parameters.

\section{Appendix}

The operator
$$
 L^{^\flat}_4=((\alpha_1x^2+1)\partial_x^2+(\alpha_2x+\alpha_3)\partial_x+\alpha_4x+\alpha_5)^2+\alpha_1\alpha_4g(g+1)x+\alpha_6
$$
commutes with an operator
$$
L^{^\flat}_{10}=P^5+\Big(\frac{5}{2} \alpha _1 (6 x \alpha _4+\alpha _2+2 \alpha _5)+3 \alpha _1^2-\frac{5 \alpha _2^2}{4}\Big)P^3+45 \alpha _1 \alpha _4 (x^2 \alpha _1+1)P^2+
$$
$$
+\frac{15}{2} \alpha _1 \alpha _4 \Big(34 x \alpha _1+3 \Big(x \alpha _2+\alpha _3\Big)\Big)P^2\partial_x-30 \alpha _1 \alpha _4 \Big(2 x^2 \alpha _1^2+\alpha _1(3 x^2 \alpha _2+6 x \alpha _3-10)-3 \alpha _2\Big)P\partial_x+
$$
$$
+q_2 P+q_1\partial_x+q_0,
$$
where
$$
P=(\alpha_1x^2+1)\partial_x^2+(\alpha_2x+\alpha_3)\partial_x+\alpha_4x+\alpha_5,
$$
$$
q_0=15 x^2 (\alpha _1+3 \alpha _2) \alpha _4^2 \alpha _1^2+\frac{3}{2} x \alpha _4 \Big(6 \alpha _1^2+(13 \alpha _2-64 \alpha _5) \alpha _1-60 \alpha _3
   \alpha _4+6 \alpha _2 (\alpha _2+20 \alpha _5)\Big) \alpha _1^2+
$$
$$
   +\frac{3}{2} \alpha _4 \Big(9 \alpha _3 \alpha _1^2+(8 \alpha _2 \alpha _3+60 \alpha _5
   \alpha _3+58 \alpha _4) \alpha _1+30 \alpha _2 \alpha _4\Big) \alpha _1,
$$
$$
q_1=15 x^2 \alpha _4 \Big(5 \alpha _1^2+4 (\alpha _2+3 \alpha _5) \alpha _1+3 \alpha _2^2\Big) \alpha _1^2-30 x \alpha _4 \Big((2 \alpha _1-3 \alpha _2)
   \alpha _3+12 \alpha _4\Big) \alpha _1^2+
$$
$$
   +15 \alpha _4 \Big(5 \alpha _1^2+2 (3 \alpha _3^2+4 \alpha _2-6 \alpha _5) \alpha _1+3 \alpha _2^2\Big) \alpha _1,
$$
$$
q_2=-135 x^2 \alpha _1^2 \alpha _4^2+\frac{15}{2} x \alpha _1 \Big(18 \alpha _1^2-2 (9 \alpha _2+14 \alpha _5) \alpha _1+\alpha _2^2\Big) \alpha _4+\frac{1}{4}
   \Big(\alpha _2^4+6 \alpha _1^3 (\alpha _2+2 \alpha _5)+
$$
$$
+\alpha _1 (-4 \alpha _2^3-8 \alpha _5 \alpha _2^2+54 \alpha _3 \alpha _4 \alpha _2+252 \alpha
   _4^2)+\alpha _1^2 (\alpha _2^2+16 \alpha _5 \alpha _2+16 \alpha _5^2-228 \alpha _3 \alpha _4)\Big).
$$

The spectral curve of $L^{^\flat}_4,L^{^\flat}_{10}$ is
$$
 w^2=z^5+c_4z^4+c_3z^3+c_2z^2+c_1z+c_0,
$$
where
$$c_4=b_4-5 \alpha _6, \ \ \  \ \ \ c_3=b_3-4b_4\alpha _6+10 \alpha _6^2, \ \ \  \ c_2=b_2-3b_3\alpha _6+6b_4\alpha _6^2-10 \alpha _6^3,
 $$
 $$
 c_1=b_1-2b_2\alpha _6+3b_3\alpha _6^2-4b_4 \alpha _6^3+5 \alpha _6^4, \  \ \ \ c_0=b_0-\alpha _6 (b_1+\alpha _6(\alpha _6(b_3-b_4 \alpha _6+\alpha _6^2)-b_2)),
$$
$$
b_4=6 \alpha_1^2+5(\alpha_2+2\alpha_5)\alpha_1-\frac{5\alpha_2^2}{2},
$$
$$
b_3=\frac{3}{16}\Big(48\alpha_1^4+96(\alpha_2+2\alpha_5)\alpha_1^3+4(-\alpha_2^2+44\alpha _5\alpha_2+44\alpha_5^2+28\alpha_3\alpha_4)\alpha_1^2-4(11\alpha_2^3+22
   \alpha_5\alpha_2^2+
$$
$$
+14\alpha_3\alpha_4\alpha_2-28\alpha_4^2)\alpha_1+11\alpha_2^4\Big),
$$

$$
b_2=\frac{1}{8}\Big(-5\alpha_2^6+3\alpha _1(10\alpha_2^3+20\alpha_5\alpha_2^2+39\alpha_3\alpha_4\alpha_2-78 \alpha _4^2)\alpha _2^2+72 \alpha _1^5(\alpha_2+2\alpha _5)+72\alpha _1^4 (\alpha _2^2+6 \alpha _5 \alpha _2+6 \alpha _5^2+
$$
$$
+8 \alpha _3 \alpha _4)+4 \alpha _1^3 \Big(-17 \alpha _2^3+120 \alpha _5^2 \alpha _2+45 \alpha _3 \alpha _4 \alpha _2+80 \alpha
   _5^3+234 \alpha _4^2+6 (\alpha _2^2+39 \alpha _3 \alpha _4) \alpha _5\Big)-
$$
$$
-3\alpha_1^2 \Big(11 \alpha _2^4+80 \alpha _5 \alpha _2^3+4 (20 \alpha _5^2+39 \alpha _3 \alpha _4) \alpha
   _2^2+12 \alpha _4 (13 \alpha _3 \alpha _5-10 \alpha _4) \alpha _2+6 \alpha _4^2 (3 \alpha _3^2-64 \alpha _5)\Big)\Big),
$$

$$
b_1=\frac{1}{16} \Big(\alpha _2^8-8 \alpha _1 (\alpha _2^3+2 \alpha _5 \alpha _2^2+9 \alpha _3 \alpha _4 \alpha _2-18 \alpha _4^2) \alpha _2^4+12 \alpha _1^5 \Big(\alpha _2^3+48 \alpha _5^2 \alpha
   _2+78 \alpha _3 \alpha _4 \alpha _2+32 \alpha _5^3+180 \alpha _4^2+
$$
$$
+6 (3 \alpha _2^2+32 \alpha _3 \alpha _4) \alpha _5\Big)+36 \alpha _1^6 \Big((\alpha _2+2 \alpha _5)^2+12 \alpha
   _3 \alpha _4\Big)+6 \alpha _1^2 \Big(3 \alpha _2^6+16 \alpha _5 \alpha _2^5+8 (2 \alpha _5^2+9 \alpha _3 \alpha _4) \alpha _2^4+
$$
$$
+12 \alpha _4 (8 \alpha _3 \alpha _5-5 \alpha _4)
   \alpha _2^3+6 \alpha _4^2 (15 \alpha _3^2-44 \alpha _5) \alpha _2^2-288 \alpha _3 \alpha _4^3 \alpha _2+288 \alpha _4^4\Big)+4 \alpha _1^3 \Big(\alpha _2^5-30 \alpha _5 \alpha _2^4-48 (2
   \alpha _5^2+
$$
$$
+3 \alpha _3 \alpha _4) \alpha _2^3-2 (32 \alpha _5^3+288 \alpha _3 \alpha _4 \alpha _5+153 \alpha _4^2) \alpha _2^2-18 \alpha _4 (27 \alpha _4 \alpha _3^2+16 \alpha _5^2
   \alpha _3-32 \alpha _4 \alpha _5) \alpha _2+36 \alpha _4^2 (-3 \alpha _5 \alpha _3^2+
$$
$$
+24 \alpha _4 \alpha _3+28 \alpha _5^2)\Big)+\alpha _1^4 \Big(-47 \alpha _2^4-160 \alpha _5 \alpha
   _2^3+96 (\alpha _5^2-6 \alpha _3 \alpha _4) \alpha _2^2+128 (4 \alpha _5^3+9 \alpha _3 \alpha _4 \alpha _5+18 \alpha _4^2) \alpha _2+8 (32 \alpha _5^4+
$$
$$
+288 \alpha _3 \alpha _4 \alpha _5^2+792 \alpha _4^2 \alpha _5+189 \alpha _3^2 \alpha _4^2)\Big)\Big),
$$

$$
b_0= \frac{3}{8} \alpha _1 \alpha _4 \Big(36 (\alpha _3 (\alpha _2+2 \alpha _5)-6 \alpha _4) \alpha _1^6+6 \Big(2 \alpha _3 \alpha _2^2+2 (6 \alpha _4+13 \alpha _3 \alpha _5) \alpha _2+32 \alpha _3 \alpha _5^2+45 \alpha _3^2 \alpha _4-48 \alpha _4 \alpha _5\Big) \alpha _1^5+
$$
$$
+\Big(-47 \alpha _3 \alpha _2^3+186 \alpha _4 \alpha _2^2-36 \alpha _3^2 \alpha _4 \alpha _2+128 \alpha
   _3 \alpha _5^3+1728 \alpha _3 \alpha _4^2+96 (\alpha _2 \alpha _3+\alpha _4) \alpha _5^2+24 \Big((15 \alpha _3^2+16 \alpha _2) \alpha _4-
$$
$$
-4 \alpha _2^2 \alpha _3\Big) \alpha _5\Big)
   \alpha _1^4+4 \Big(\alpha _3 \alpha _2^4-(29 \alpha _4+24 \alpha _3 \alpha _5) \alpha _2^3-3 (15 \alpha _4 \alpha _3^2+16 \alpha _5^2 \alpha _3+2 \alpha _4 \alpha _5) \alpha _2^2-4
   \Big(18 \alpha _4 \alpha _5 \alpha _3^2+(4 \alpha _5^3+
$$
$$
+63 \alpha _4^2) \alpha _3-12 \alpha _4 \alpha _5^2\Big) \alpha _2+2 \alpha _4 \Big(16 \alpha _5^3+180 \alpha _3 \alpha _4 \alpha _5-27
   (\alpha _3^3-8 \alpha _4) \alpha _4\Big)\Big) \alpha _1^3+6 \Big(3 \alpha _3 \alpha _2^5+12 \alpha _3 \alpha _5 \alpha _2^4+
$$
$$
+2 (9 \alpha _4 \alpha _3^2+4 \alpha _5^2 \alpha _3-8 \alpha _4
   \alpha _5) \alpha _2^3-4 \alpha _4 (-3 \alpha _5 \alpha _3^2+3 \alpha _4 \alpha _3+4 \alpha _5^2) \alpha _2^2-6 \alpha _4^2 (-3 \alpha _3^3+20 \alpha _5 \alpha _3+8 \alpha _4)
   \alpha _2+
$$
$$
+12 \alpha _4^3 (16 \alpha _5-3 \alpha _3^2)\Big) \alpha _1^2-2 \alpha _2^2 \Big(4 \alpha _3 \alpha _2^4-6 (\alpha _4-\alpha _3 \alpha _5) \alpha _2^3+3 \alpha _4 (3
   \alpha _3^2-4 \alpha _5) \alpha _2^2-36 \alpha _3 \alpha _4^2 \alpha _2+36 \alpha _4^3\Big) \alpha _1+
$$
$$
+\alpha _2^6 (\alpha _2 \alpha _3-2 \alpha _4)\Big).
$$

\end{document}